# Progress Towards Opto-Electronic Characterization of Indium Phosphide Nanowire Transistors at milli-Kelvin temperatures

L.H. Willems van Beveren[1*], J.C. McCallum[1], H.H. Tan[2] and C. Jagadish[2]

[1]Centre for Quantum Computation and Communication Technology (CQC²T),
School of Physics, University of Melbourne, VIC 3010, Australia
[2]Research School of Physics and Engineering, Australian National University, ACT 0200, Australia
*Corresponding author: Email laurensw@unimelb.edu.au

**Abstract:** In this paper we present our progress towards the opto-electronic characterization of indium phosphide (InP) nanowire transistors at milli-Kelvin (mK) temperatures. First, we have investigated the electronic transport of the InP nanowires by current-voltage (I-V) spectroscopy as a function of temperature from 300 K down to 40 K. Second, we show the successful operation of a red light emitting diode (LED) at liquid-Helium (and base) temperature to be used for opto-electronic device characterization.

**1 Introduction:** Semiconductor nanowires offer the possibility to fabricate nanoscale devices that exhibit enhanced functionality compared to conventional device structures, opening up new opportunities in fields including electronics, quantum computing, optoelectronics and sensing [1].

Nanowires can be formed using a variety of materials including elemental semiconductors and metals, compound semiconductors, heterostructures, carbon nanotubes and conducting polymers. Due to their small scale, defect-free heterostructure nanowires can be formed that are otherwise not favorable. The high surface to volume ratio results in their electronic properties being strongly influenced by external factors making them highly sensitive sensors [2].

The large variety of nanowire materials and structures available offer many possibilities for development of new devices. Substantial progress has been made in terms of synthesis, material characterization and application of nanowires into functional devices such as photon sources and sensors.

However, there remain unanswered questions regarding the role of defects, interfaces and electronic states on the performance of semiconductor nanowire devices and in functionalizing nanowires for bio-sensing applications. Some of these questions can be answered by investigating the electronic transport properties of nanowire-based device architectures at milli-Kelvin temperatures.

**2 I-V characteristic of an InP nanowire:** Here, we report the current-voltage (I-V) characteristic of a suspended InP nanowire connected to two gold (Au) electrodes as a function of temperature. The nanowires were synthesized by a chemical vapour deposition (CVD) process [3], then transferred (ultrasonic agitation) onto a pre-patterned Si/SiO$_2$ substrate and electrically contacted to electrodes by focused ion-beam (FIB) based direct write technology which deposits platinum (Pt).

A scanning electron micrograph (SEM) of the typically 50 nm wide (in diameter) suspended InP nanowire transistor is shown in Fig. 1.

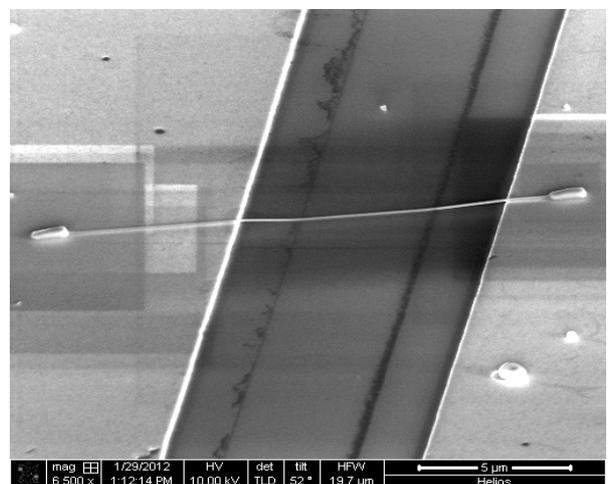

Fig. 1. The about 10 μm long suspended InP nanowire in between two gold electrodes.

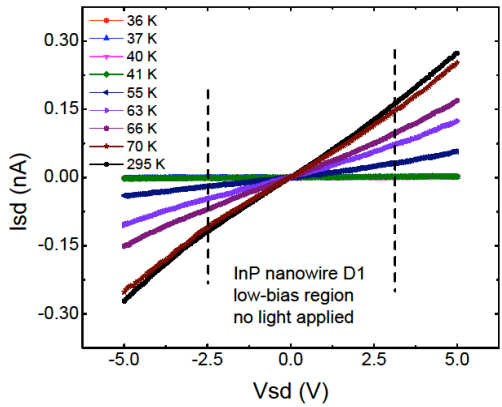

Fig. 2. The I-V characteristic of the InP nanowire as a function of temperature.

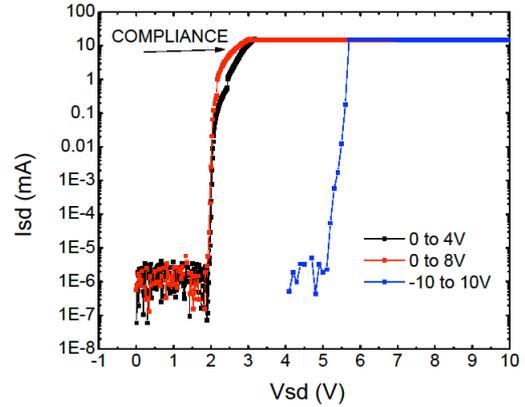

Fig. 3. The I-V characteristic of a red LED at liquid Helium temperature (4 K).

In Fig. 2 we show the I-V characteristic of the intentionally undoped InP nanowire (in the dark) as a function of temperature from room temperature down to 40 K. The resistance of the nanowire becomes unmeasurable for lower temperatures [4]. Carrier freeze out occurs.

## 3 Red LED at liquid-He temperatures:
We want to use a red low-temperature mounted light-emitting-diode (LED) to excite by photo-illumination extra charge carriers in the InP nanowire below 40 K.

To demonstrate the red LED operates at such low temperatures, we have measured the I-V characteristic of a red LED (4 K) which can be seen in Fig. 3. The voltage source reaches compliance for a current level of ~15 mA.

Note how the turn-on point of the I-V curve moves to lower voltage (red/black are trace 2 and 3) for subsequent sweeps (blue is first trace). This is due to power dissipation ($I^2 \ast R$ heating) when too large currents (>10 mA) are sent through the LED and any series resistance that may be present (wiring / RC filters) in the fridge. We have also measured this LED at a 12 mK base temperature and the 'turn-on' for this LED corresponds to a > 6 V bias voltage across the *p-n* junction.

Note that this particular LED has not been thermally cycled [5]. The I-V data at 4K (when power dissipation occurs – red/black trace) is comparable to room temperature (RT) and liquid nitrogen temperature I-V characteristics. This implies that the LED can warm up significantly during operation and current limits have to be adhered to in order to keep the device under test (and LED) cold.

In general, a good way to test if an LED will work at liquid helium temperatures is to cool it down to liquid $N_2$ temperature and see if the I-V corresponds to the RT data. Other LEDs (HP/Agilent and Siemens) did show a large difference in I-V as a function of temperature and do not emit any visible light at liquid helium temperatures. In the future we will calibrate the light emission of this red LED.

### Acknowledgements

This work has been supported by a University of Melbourne Materials Institute (MMI) interdisciplinary (ID) seed-funding grant (2012). The authors acknowledge P. Prasai for device fabrication assistance.

### References

[1] Y. Li, F. Qian, J. Xiang, and C.M. Lieber, "Nanowire electronic and optoelectronic devices", Materials Today 9, pp. 18-y Month 2006.

[2] T.C. Nguyen, W. Qiu, M. Altissimo, et al., "CMOS compatible nanowire biosensors", in "Nanofabrication: materials and techniques", CRC Press 2012.

[3] Q. Gao, H.H. Tan, et al., "Growth and properties of III–V compound semiconductor heterostructure nanowires", Semicond. Sci. Technol. 26, pp. 014035 x-y Month 2011.

[4] D. Adachi, in "Properties of Group-IV, III–V and II–VI Semiconductors", John Wiley & Sons, Ltd. 2005.

[5] H. Akimoto, A. Marchenkov, et al., "Forward current–voltage characteristics of light-emitting diodes at low temperatures", Cryogenics 38, pp. 451-y Month 1998.